\documentclass[twocolumn,preprintnumbers,aps,prb,floatfix,showpacs,superscriptaddress]{revtex4}
\usepackage{graphicx}
\usepackage{bm}
\usepackage{latexsym,color}

\begin{document}

\title{Phase formation characteristics and magnetic properties of bulk Ni$_{2}$MnGe Heusler alloy}
\author{U. Adem}
\affiliation{Department of Engineering Physics, Faculty of Engineering, Ankara University, 06100 Besevler, Ankara, Turkey}
\author{\.{I}. Din\c{c}er}
\affiliation{Department of Engineering Physics, Faculty of Engineering, Ankara University, 06100 Besevler, Ankara, Turkey}

\author{S. Akt\"{u}rk}
\affiliation{Department of Physics, Faculty of Sciences and Letters, Mu\u{g}la University, 48000 K\"{o}tekli, Mu\u{g}la, Turkey}
\author{M. Acet}
\affiliation{Experimental Physics, Duisburg-Essen University, 47048 Duisburg, German}

\author{Y. Elerman}
\affiliation{Department of Engineering Physics, Faculty of Engineering, Ankara University, 06100 Besevler, Ankara, Turkey}

\date{\today}

\begin{abstract}

We have systemically studied the effects of annealing temperature and alloy composition  on the structural and magnetic
properties of bulk Ni$_{2}$MnGe and Ni$_{2.1}$Mn$_{0.9}$Ge Heusler alloys. We have observed that both annealing temperature and the alloy composition drastically alter the phases found in the samples due to the presence of competing ternary phases. Annealing at 900 and 950 $^{\circ}$C for both alloy compositions facilitate the formation of L2$_{1}$ Heusler phase. Nevertheless, formation of Ni$_{5}$Mn$_{4}$Ge$_{3}$ and Ni$_{16}$Mn$_{6}$Ge$_{7}$ phases cannot be prevented for Ni$_{2}$MnGe and Ni$_{2.1}$Mn$_{0.9}$Ge alloys, respectively. In order to estimate the magnetic contribution of the Ni$_{5}$Mn$_{4}$Ge$_{3}$ impurity phase to that of the parent Ni$_{2}$MnGe, we have also synthesized pure Ni$_{5}$Mn$_{4}$Ge$_{3}$ alloy. Antiferromagnetic nature of Ni$_{5}$Mn$_{4}$Ge$_{3}$ with low magnetization response allows us to reveal the magnetic response of the stoichiometric bulk Ni$_{2}$MnGe. Bulk Ni$_{2}$MnGe shows simple ferromagnetic behavior with a Curie temperature of 300 K, in agreement with the previous results on thin films. Despite the divergence of magnetization curves between field cooled (FC) and field heated (FH) modes, stoichiometric Ni$_{2}$MnGe alloy does not undergo a martensitic phase transition based on our variable temperature x-ray diffraction experiments.

\end{abstract}

\maketitle

\newpage

Ni-Mn based Heusler alloys constitute a prime family of compounds that have been studied extensively due to the presence of a magnetostructural phase transition which makes them useful for magnetic refrigeration and magnetic shape memory applications such as actuators\cite{Acet, Gutfleisch, Ullakko}. Upon cooling, Ni-Mn-Z (where Z = Ga,In,Sn,Sb) undergo a phase transition into a martensitic phase. Only for Ni$_{2}$MnGa phase, martensitic phase transition occurs for the stoichiometric Heusler composition in the ferromagnetic state. Associated with the first order martensitic phase transition, (magnetic) shape memory effects are observed in these alloys. Magnetic and structural phase transition temperatures can be tailored to yield large caloric effects, which makes them useful for magnetic refrigeration applications\cite{Acet}.

Despite the abundance of studies on other Heusler compositions, there have been relatively few studies on Ni$_{2}$MnGe both in bulk and thin film form. Lund et al. have reported the growth of thin films of Ni$_{2}$MnGe using Molecular Beam Epitaxy (MBE) and obtained ferromagnetic films with a Curie temperature of 300 K \cite{Lund}. Kim et al. have grown Ni$_{2}$MnGe films using flash evaporation. Their films were ferromagnetic with a Curie temperature of 280 K \cite{Kim}. Only Oksenenko et al. have reported the structural and magnetic properties of stoichiometric bulk Heusler composition Ni$_{2}$MnGe\cite{Oksenenko}. They could not obtain a pure Heusler phase and observed complex magnetization behaviour accordingly. Si et al. synthesized bulk Ni$_{2.1}$Mn$_{0.9}$Ge and  Ni$_{2.2}$Mn$_{0.8}$Ge\cite{Si}. They claimed to obtain pure off-stoichiometric Heusler phase. Their magnetic measurements show a simple ferromagnetic behavior with an ordering temperature T$_{C}$ of 246 K for Ni$_{2.1}$Mn$_{0.9}$Ge and 151 K for Ni$_{2.2}$Mn$_{0.8}$Ge.

The key property that renders other Ni-Mn based Heusler alloys useful for applications is the presence of a first-order martensitic phase transition from high temperature L2$_{1}$ phase to a low temperature martensite phase\cite{Acet}. Zayak et al. predicted an instability of L2$_{1}$ structure in Ni$_{2}$MnGe alloys using first-principles calculations but this has not been observed so far\cite{Zayak}. Recently, there has been another prediction\cite{Luo}, suggesting the likelihood of a transition from the cubic austenite phase to a tetragonal martensite phase. We have undertaken a systematic study to clarify the structural and magnetic properties of stoichiometric bulk Ni$_{2}$MnGe and to address the martensitic phase transition predictions. In addition to the stoichiometric composition, we have also synthesized Ni$_{2.1}$Mn$_{0.9}$Ge alloy, in order to reproduce the pure L2$_{1}$ Heusler phase claimed in ref.\cite{Si}. Another off-stoichiometric sample with the determined composition of Ni$_{1.91}$Mn$_{0.94}$Ge$_{1.15}$ was also used to study the effect of Ni off-stoichiometry and demonstrate the effect of annealing temperature. We have observed that annealing temperature and alloy composition significantly affect the phases encountered in the alloys. Finally, we show the lack of a martensitic phase transition using variable temperature x-ray diffraction experiments.

\section{Experimental}

Ni$_{2}$MnGe, Ni$_{2.1}$Mn$_{0.9}$Ge, Ni$_{5}$Mn$_{4}$Ge$_{3}$ and Ni$_{1.91}$Mn$_{0.94}$Ge$_{1.15}$ alloys were synthesized by arc-melting the constituting elements under Ar atmosphere using an Arc Melter. The alloys were melted 5 times to assure melt homogeneity. Ni$_{2}$MnGe and Ni$_{2.1}$Mn$_{0.9}$Ge alloys were annealed first at 900$^{\circ}$C for 3 days in order to reproduce the synthesis conditions in ref.\cite{Si} and then subsequently annealed at 950 $^{\circ}$C for 4 days to see the effects of higher temperature on the stability of phases. Ni$_{5}$Mn$_{4}$Ge$_{3}$ alloy was annealed at 800$^{\circ}$C for 6 days following ref.\cite{Gladyshevskii}. Ni$_{1.91}$Mn$_{0.94}$Ge$_{1.15}$ was annealed at  800$^{\circ}$C for a week and then subsequently annealed at 950$^{\circ}$C for another week. Annealing was done with samples closed in quartz tubes under Ar atmosphere and alloys were quenched in ice water. Composition of the alloys were checked using Energy Dispersive X-ray Spectroscopy(EDS) with a Zeiss EVO 40 Scanning Electron Microscope (SEM). Crystal structure of the alloys were determined using x-ray diffraction experiments with a Rigaku D-Max 2200 diffractometer with Mo K-alpha radiation at room temperature. Variable temperature x-ray diffraction experiments were carried out using a Rigaku Smartlab diffractometer having a Cu K-alpha source equipped with a commercial temperature attachment, between 83-573 K. Le Bail fits to the x-ray diffraction data were done using GSAS software package\cite{Larson}. Magnetization of the samples were measured in a MPMS magnetometer between 10-380 K.

\section{Results and Discussion}

Fig. 1 shows the Le Bail fit of the x-ray diffraction data of the stoichiometric Ni$_{2}$MnGe alloy annealed at 900 $^{\circ}$C for 3 days collected at room temperature. All peaks can be indexed with Heusler L2$_{1}$ phase with the space group Fm\={3}m and Ni$_{5}$Mn$_{4}$Ge$_{3}$ impurity phase.  Ni$_{5}$Mn$_{4}$Ge$_{3}$ is reported to crystallize in hexagonal P6$_{3}$/mmc space group \cite{Gladyshevskii}. Lattice parameters from the Le Bail fit were obtained as \emph{a} = 5.8224(4){\AA} for the cubic Ni$_{2}$MnGe phase and \emph{a} = 4.9288(9){\AA} and \emph{c} = 7.680(3){\AA} for the Ni$_{5}$Mn$_{4}$Ge$_{3}$ phase, respectively. Upon further annealing at 950 $^{\circ}$C for 4 days, Ni$_{5}$Mn$_{4}$Ge$_{3}$ peaks lose their intensity (shown in the inset of Fig. 1), however a pure Ni$_{2}$MnGe phase cannot be obtained.

\begin{figure}[tbp]
\centering
\includegraphics[width=\columnwidth]{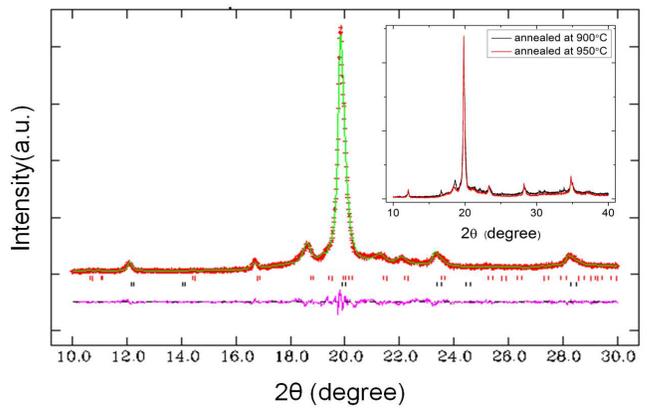}
\caption{X-ray diffraction pattern of Ni$_{2}$MnGe sample annealed at 900 $^{\circ}$C for 3 days. Black ticks denote L2$_{1}$ Heusler peaks whereas red ticks show Ni$_{5}$Mn$_{4}$Ge$_{3}$ impurity phase. Observed data is red, calculated is in green and the difference is in pink. The inset shows a comparision of the diffraction patterns corresponding to the sample annealead at 900 $^{\circ}$C and subsequently at 950 $^{\circ}$C.}\label{fig1}
\end{figure}

In Fig. 2(a), Backscattered Electron (BSD)image of Ni$_{2}$MnGe alloy is shown. There are thin dark regions within the lightly colored matrix. Point analysis from dark regions show a stoichiometry close to Ni$_{5}$Mn$_{4}$Ge$_{3}$ in agreement with the XRD data. Neighbouring lightly colored regions show stoichiometric Ni$_{2}$MnGe composition (Table 1).

\begin{figure}[tbp]
\centering
\includegraphics[width=0.9\columnwidth]{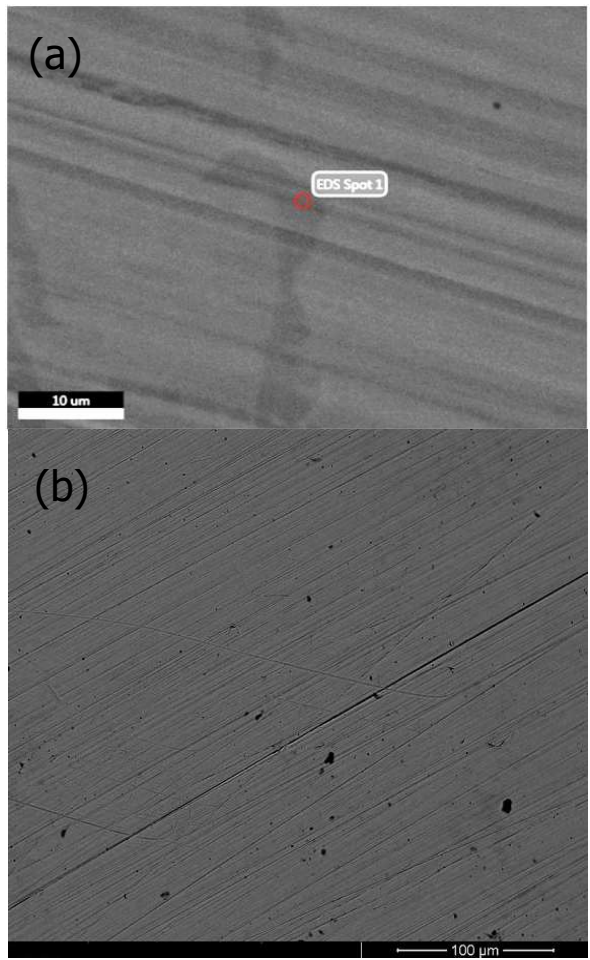}
\caption{Backscattered Electron (BSD) SEM images of (a)Ni$_{2}$MnGe alloy, annealed at 900 $^{\circ}$C for 3 days. Figure shows a zoomed-in region where dark colored Ni$_{5}$Mn$_{4}$Ge$_{3}$ phase can be seen, surrounded by the lighter colored regions of Ni$_{2}$MnGe phase, (b) Ni$_{2.1}$Mn$_{0.9}$Ge alloy, annealed at 900 $^{\circ}$C for 3 days.}\label{fig2}
\end{figure}

\begin{table}[htb]
\centering
\begin{ruledtabular}
\begin{tabular}{|c|c|c|c|c|}

    \hline

 Composition & Ni & Mn & Ge  \\
  \hline
  Ni$_{2}$MnGe (dark regions) & 45.83 (41.7)  & 31.45 (33.3)    & 22.72 (25) \\
  Ni$_{2}$MnGe (light regions) & 49.83 (50)  & 25.42 (25)    & 24.75 (25) \\
  Ni$_{2.1}$Mn$_{0.9}$Ge      & 51.91 (52.5)  & 22.81 (22.5)   & 25.27 (25)   \\
  Ni$_{1.91}$Mn$_{0.94}$Ge$_{1.15}$      & 47.8 (47.8)  & 23.5 (23.5)   & 28.7 (28.7)   \\
  \hline
\end{tabular}
\caption{Elemental analysis results of Ni$_{2}$MnGe(dark regions), Ni$_{2}$MnGe(light regions), Ni$_{2.1}$Mn$_{0.9}$Ge and Ni$_{1.91}$Mn$_{0.94}$Ge$_{1.15}$. The numbers in parenthesis denote the expected values for the respective phases: i.e. Ni$_{5}$Mn$_{4}$Ge$_{3}$ stoichiometry for Ni$_{2}$MnGe (dark regions), Ni$_{2}$MnGe stoichiometry for Ni$_{2}$MnGe (light regions) and Ni$_{2.1}$Mn$_{0.9}$Ge stoichiometry for Ni$_{2.1}$Mn$_{0.9}$Ge.}
%\label{table1}
\end{ruledtabular}
\end{table}

Le Bail fit of the x-ray diffraction data of Ni$_{2.1}$Mn$_{0.9}$Ge annealed at 900 $^{\circ}$C shown in Fig. 3 can be indexed using a mixture of L2$_{1}$ Ni$_{2}$MnGe Heusler phase together with Mn$_{6}$Ni$_{16}$Ge$_{7}$. Mn$_{6}$Ni$_{16}$Ge$_{7}$ phase is cubic with the space group Fm\={3}m\cite{Gladyshevskii}. Refined lattice parameters were \emph{a} = 5.7881(5){{\AA}} and \emph{a} = 11.4164(6){{\AA}} for Ni$_{2}$MnGe and  Mn$_{6}$Ni$_{16}$Ge$_{7}$, respectively. Further annealing at 950 $^{\circ}$C doesn't change the stability of the phases.

In Fig. 2(b), BSD image of Ni$_{2.1}$Mn$_{0.9}$Ge alloy is shown. Unlike the case in Ni$_{2}$MnGe composition, no dark and lightly colored regions could be observed, therefore the presence of a second phase with Mn$_{6}$Ni$_{16}$Ge$_{7}$ stoichiometry couldn't be verified. On the other hand, elemental analysis from a large area showed that the target composition has been obtained (Table 1).

\begin{figure}[tbp]
\centering
\includegraphics[width=\columnwidth]{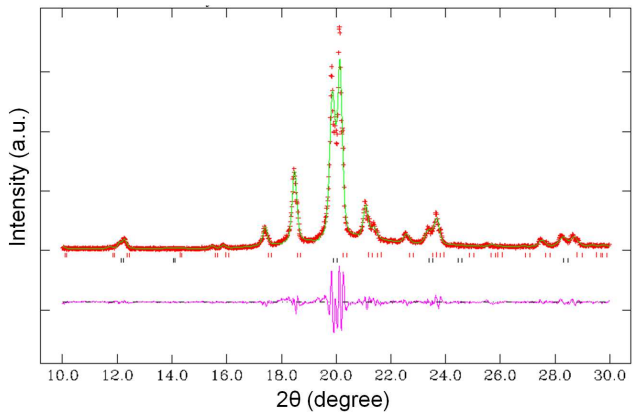}
\caption{Le Bail fit of the x-ray diffraction pattern of Ni$_{2.1}$Mn$_{0.9}$Ge sample annealed at 900 $^{\circ}$C for 3 days. Black ticks denote L2$_{1}$ Heusler peaks whereas red ticks show Ni$_{16}$Mn$_{6}$Ge$_{7}$ impurity phase.}\label{fig3}
\end{figure}

In Fig. 4, we show the temperature dependence of magnetization of Ni$_{2}$MnGe, Ni$_{2.1}$Mn$_{0.9}$Ge and Ni$_{5}$Mn$_{4}$Ge$_{3}$ samples measured at 50 Oe. Sharp increase in magnetization in both FC and FH modes define the ferromagnetic Curie temperature as 300 K for Ni$_{2}$MnGe and 320 K for Ni$_{2.1}$Mn$_{0.9}$Ge. There is divergence between FC and FH modes, hinting the possibility of the presence of a first order martensitic phase transition. In order to quantify the contribution of Ni$_{5}$Mn$_{4}$Ge$_{3}$ impurity to the magnetic behaviour of Ni$_{2}$MnGe, a pristine Ni$_{5}$Mn$_{4}$Ge$_{3}$ alloy was synthesized. Le Bail fit of XRD data of Ni$_{5}$Mn$_{4}$Ge$_{3}$ alloy is shown in Fig. 5. Refined lattice parameters are \emph{a} = 4.8865(3){{\AA}} and \emph{c} = 7.6662(7){{{\AA}}}. Temperature dependence of magnetization of pure Ni$_{5}$Mn$_{4}$Ge$_{3}$ is shown in the inset of Fig. 4. Ni$_{5}$Mn$_{4}$Ge$_{3}$ shows complex magnetic response, dominantly antiferromagnetic, however since the magnitude of magnetization is very small, magnetic contribution of Ni$_{5}$Mn$_{4}$Ge$_{3}$ to the behavior of parent Ni$_{2}$MnGe is negligible.

\begin{figure}[tbp]
\centering
\includegraphics[width=\columnwidth]{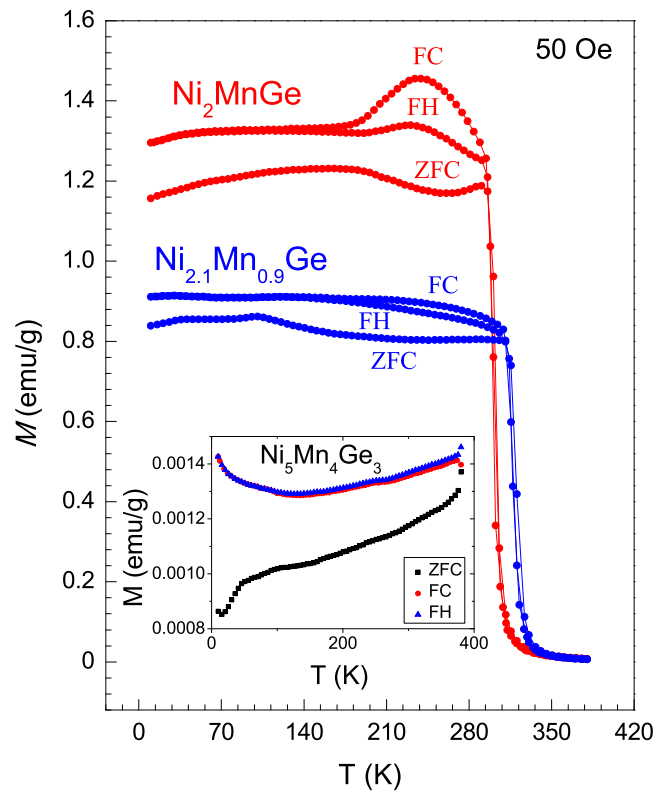}
\caption{Temperature dependence of magnetization of Ni$_{2}$MnGe, Ni$_{2.1}$Mn$_{0.9}$Ge and Ni$_{5}$Mn$_{4}$Ge$_{3}$ measured at 50 Oe.  }\label{fig4}
\end{figure}

\begin{figure}[tbp]
\centering
\includegraphics[width=\columnwidth]{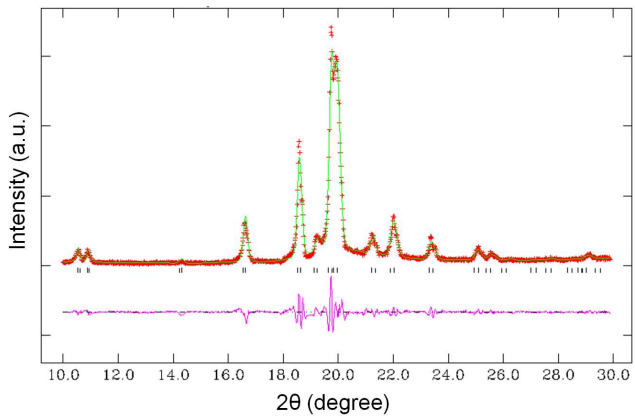}
\caption{Le Bail fit of x-ray diffraction pattern for  Ni$_{5}$Mn$_{4}$Ge$_{3}$ sample. All peaks correspond to Ni$_{5}$Mn$_{4}$Ge$_{3}$.}
\label{fig5}
\end{figure}

Our studies show that Ni$_{2.1}$Mn$_{0.9}$Ge alloy contains two phases: L2$_{1}$ Ni$_{2}$MnGe Huesler phase and Mn$_{6}$Ni$_{16}$Ge$_{7}$ (Fig. 3). Si et al. however claim that they observe only L2$_{1}$ Heusler phase\cite{Si}. Magnetic properties of Mn$_{6}$Ni$_{16}$Ge$_{7}$ are unknown but provided that it shows dominantly antiferromagnetic properties similar to Ni$_{5}$Mn$_{4}$Ge$_{3}$, simple ferromagnetic behaviour that Si et al. have reported, assigned to L2$_{1}$ Heusler phase, can be understood.

Observation of three ternary phases (Ni$_{2}$MnGe, Mn$_{6}$Ni$_{16}$Si$_{7}$ and Ni$_{5}$Mn$_{4}$Ge$_{3}$) in a narrow composition range might seem unusual when one considers the number of ternary phases in Ni-Mn-Ga(only one)\cite{raey} however formation of quite a few ternary phases are known for both Ni-Mn-Ge\cite{Gladyshevskii} and Ni-Mn-Si systems\cite{raey2}. Ni-Mn-Si ternary phase diagram hosts many ternary phases, including Mn$_{6}$Ni$_{16}$Si$_{7}$ \cite{raey2, kolenda}. In the following, we will show that annealing temperature plays a crucial role for the formation of Heusler phase. We have observed a very similar magnetic behavior to that in Oksenenko et al., in a Mn and Ni deficient composition, Ni$_{1.91}$Mn$_{0.94}$Ge$_{1.15}$, when the sample was annealed at 800$^{\circ}$C, not at 900 or 950 $^{\circ}$C. Temperature dependence of magnetization of this alloy is shown in Fig. 6(a). There are multiple anomalies, which probably correspond to different magnetic ordering phenomena. The different and complex magnetic behavior that we observed can be ascribed to the different phases present in this alloy. Different to the Ni$_{2}$MnGe and Ni$_{2.1}$Mn$_{0.9}$Ge compositions, Ni$_{1.91}$Mn$_{0.94}$Ge$_{1.15}$ annealed at 800$^{\circ}$C doesn't contain L2$_{1}$ Heusler phase but includes Mn$_{6}$Ni$_{16}$Ge$_{7}$ and Ni$_{5}$Mn$_{4}$Ge$_{3}$ phases. In Fig. 7, Le Bail fit of the X-ray diffraction pattern of Ni$_{1.91}$Mn$_{0.94}$Ge$_{1.15}$ annealed at 800 $^{\circ}$C is shown. Black tickmarks denote peaks belonging to Ni$_{5}$Mn$_{4}$Ge$_{3}$ phase, with the refined lattice parameters \emph{a} = 4.8944(3){{{\AA}}} and \emph{c} =  7.683(2){{{\AA}}}, while red tickmarks correspond to Mn$_{6}$Ni$_{16}$Ge$_{7}$ phase, with the refined cubic lattice parameter \emph{a} = 11.526(1){{{\AA}}}. More experiments are necessary to understand the magnetic anomalies in this alloy.

Annealing Ni$_{1.91}$Mn$_{0.94}$Ge$_{1.15}$ composition further at 950 $^{\circ}$C yields a mixture of L2$_{1}$ Heusler Ni$_{2}$MnGe phase with Mn$_{6}$Ni$_{16}$Ge$_{7}$ present as the second phase (Fig. 7). Refined lattice parameters were \emph{a} = 5.8115(2){{\AA}} for the Ni$_{2}$MnGe phase and \emph{a} = 11.475(2){{\AA}} for Mn$_{6}$Ni$_{16}$Ge$_{7}$. Temperature dependence of magnetization of this alloy annealed at 950$^{\circ}$C is given in Fig. 6(b). Due to dominant presence of cubic Heusler phase, magnetization shows ferromagnetic ordering around 315 K, similar to the stoichiometric Ni$_{2}$MnGe alloy as well as to off-stoichiometric Ni$_{2.1}$Mn$_{0.9}$Ge, both containing cubic Heusler phases, as shown in the Fig. 4.

\begin{figure}[tbp]
\centering
\includegraphics[width=\columnwidth]{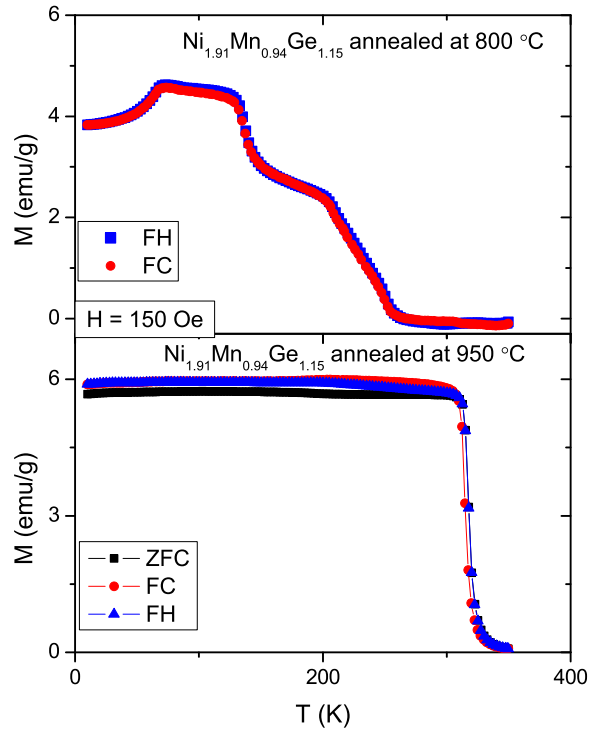}
\caption{Temperature dependence of magnetization of Ni$_{1.91}$Mn$_{0.94}$Ge$_{1.15}$ sample (a) annealed at 800$^{\circ}$C, (b)  annealed at 950$^{\circ}$C, both measured at 150 Oe.}
\protect\label{Fig. 6}
\end{figure}

\begin{figure}[tbp]
\centering
\includegraphics[width=\columnwidth]{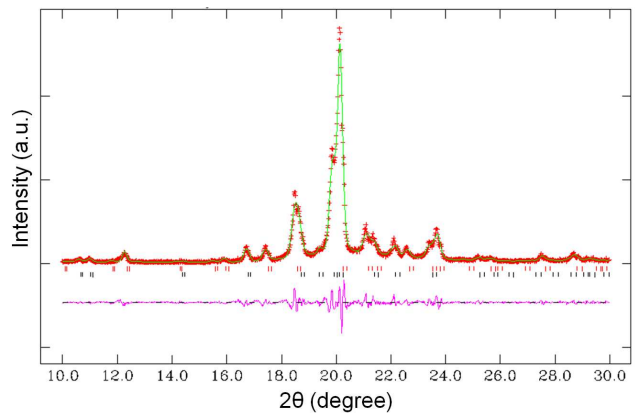}
\caption{Le Bail fit of the x-ray diffraction data for Ni$_{1.91}$Mn$_{0.94}$Ge$_{1.15}$ annealed at 800$^{\circ}$C. Black tickmarks denote Ni$_{5}$Mn$_{4}$Ge$_{3}$ peaks, while red tickmarks correspond to Mn$_{6}$Ni$_{16}$Ge$_{7}$ phase.}
\label{Fig. 7}
\end{figure}

\begin{figure}[tbp]
\centering
\includegraphics[width=\columnwidth]{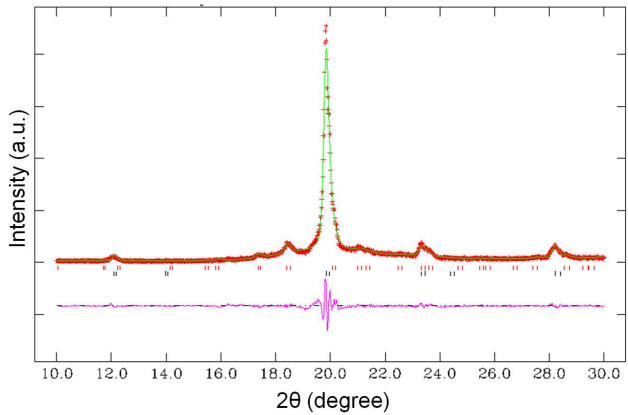}
\caption{Le Bail fit of the x-ray diffraction data for Ni$_{1.91}$Mn$_{0.94}$Ge$_{1.15}$ subsequently annealed at 950$^{\circ}$C after annealing at 800$^{\circ}$C. Black tickmarks denote L2$_{1}$ Ni$_{2}$MnGe peaks, while red tickmarks correspond to Mn$_{6}$Ni$_{16}$Ge$_{7}$ phase. \protect\label{Fig. 8} }
\end{figure}

Another major outcome of our study is the observation of absence of a martensitic phase transition. As it has been discussed in the introduction section, there are theoretical calculations claiming the presence a structural instability in stoichiometric Ni$_{2}$MnGe that can result in a martensitic phase transition\cite{Zayak, Luo}. We are able to assess this prediction satisfactorily by identifying the phases present in our samples. We have used Ni$_{1.91}$Mn$_{0.94}$Ge$_{1.15}$ sample annealed at 950$^{\circ}$C for this purpose. Refined diffraction pattern collected at RT was shown in Fig. 8. X-ray diffraction spectra of Ni$_{1.91}$Mn$_{0.94}$Ge$_{1.15}$ sample, collected at 83 K, RT, 423 K and 573 K are shown in Fig. 9. Throughout this temperature range, including the part where a divergence between FC and FH curves in magnetization was observed, both cubic Heusler peaks and peaks belonging to Mn$_{6}$Ni$_{16}$Ge$_{7}$ phase have been observed, with no additional peaks appearing. Therefore, we conclude that no martensitic transiton takes place for the stoichiometric Ni$_{2}$MnGe alloy. An additional reflection corresponding to (420) plane, belonging to the cubic Heusler phase is visible at all temperatures, different than the data collected at RT with Mo-source lab x-ray diffractometer. This might be due to the different data collection times. The lack of a structural martensitic phase transition is surprising. Typically, a divergence in magnetization between FC and FH modes of magnetization originates from a structural phase transition. Lack of a structural phase transition in our case suggests a magnetic origin for the divergence between FC and FH modes. Neutron diffraction experiments are necessary to reveal the origin of the divergence between FC and FH modes. The lack of martensitic phase transition is supporting the previous results on Ge substitution in Ni$_{2}$MnGa$_{1-x}$Ge$_{x}$\cite{Gos}. For x = 0.4, it was reported that martensitic transition temperature decreased from 200 K to 50 K compared to the Ni$_{2}$MnGa, suggesting the suppression effect of Ge on the martensitic phase transition.

\begin{figure}[tbp]
\centering
\includegraphics[width=\columnwidth]{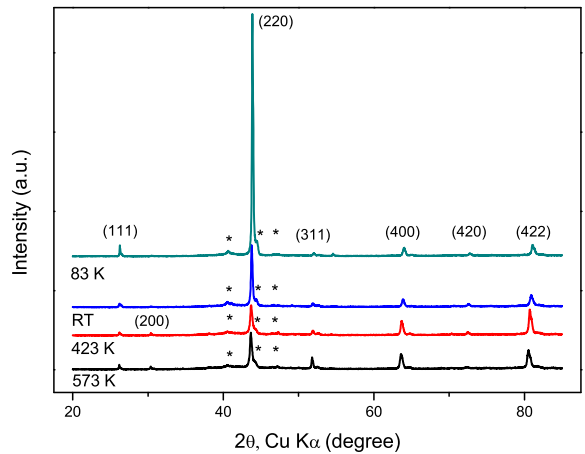}
\caption{X-ray diffraction spectrum of Ni$_{1.91}$Mn$_{0.94}$Ge$_{1.15}$ sample annealed at 950$^{\circ}$C for 6 days, collected at 83 K, RT, 423 K and 573 K. Star symbols denote the reflections belonging to the Mn$_{6}$Ni$_{16}$Ge$_{7}$ phase; the rest of the reflections belong to the parent Ni$_{2}$MnGe phase.}\label{Fig. 9}
\end{figure}

\section{Conclusions}

In conclusion, we synthesized alloy compositions close to the stoichiometric Heusler Ni$_{2}$MnGe and observed that in a narrow composition range, there are two other thermodynamically stable phases in addition to the L2$_{1}$ Ni$_{2}$MnGe structure: Ni$_{5}$Mn$_{4}$Ge$_{3}$ and Mn$_{6}$Ni$_{16}$Ge$_{7}$. By estimating the contribution from these phases, we have measured the intrinsic magnetic response of bulk Ni$_{2}$MnGe, addressing the inconsistency in previous reports. This also allowed us to evaluate the presence of a martensitic phase transition. We conclude from our variable temperature x-ray diffraction measurements that no martensitic phase transition takes place at the stoichiometric Heusler composition Ni$_{2}$MnGe. Due to the presence of various thermodynamically stable phases in equilibrium, forcing a martensitic phase transition by moving to an off-stoichiometric composition would be unlikely.

\begin{acknowledgments}

U. A. acknowledges E. Duman for helpful discussions and the financial support of Turkish Scientific and Technological Research Council (T\"{U}B\.{I}TAK) via 2232 Programme.

\end{acknowledgments}

\end{document}